# A modified micromorphic model based on micromechanics for granular materials

Chenxi Xiu[a], Xihua Chu[a,*]

[a] School of Civil Engineering, Wuhan University, Wuhan, 430072, China

**\*** Corresponding author.
Tel.: +0086-27-68772230
E-mail address: **Chuxh@whu.edu.cn**

**Abstract**：The purpose of this study is to propose a modified micromorphic continuum model for granular materials based on a micromechanics approach. In this model, Cauchy stress and the couple stress are symmetric conjugated with the symmetric strain and the symmetric curvature respectively, and the relative stress measures are asymmetric conjugated with the asymmetric relative strain measures. This modified micromorphic model considers a continuum material point as a granular volume element whose deformation behavior is influenced by the translation and the rotation of particles. And this study proposes that the microscopic actual motion is decomposed into a macroscopic motion and a fluctuation between the macro-micro motion. Based on this decomposition, the micromorphic constitutive relationships are derived for granular materials. In the constitutive relationships, the macroscopic constitutive relationships are first-order because of the introduce of the independent rotation of particle instead of the second-order micro-deformation gradient. Furthermore, the complex constitutive moduli in the micromorphic model are obtained in the expressions of the microstructural information such as the contact stiffness and the internal length.

## 1. Introduction

Granular materials are composed of solid particles and voids between particles with a high degree of heterogeneity and complex mechanical behaviors such as multi-scale behaviors, crushability, anisotropy [1-5]. These macroscopic behaviors of granular materials are closely related to the microstructure and interaction between particles. Chang and Ma [4] believed that a microstructural continuum approach can consider the effect of the microstructure and the interaction. The microstructural continuum approach can describe the macroscopic measures reflecting discrete characteristics, and then develops the macroscopic constitutive relationship. This approach has been widely applied in study on mechanical behaviors of granular materials [4,6-11]. The key of the approach is how to establish a relationship between micromechanical responses and macroscopic continuous behaviors for granular materials [12]. Considering the discrete characteristics of granular materials, definitions of the strain and the stress based on classical continuum are no longer applicable. Therefore, studies [12-17] are devoted to the development of microstructural descriptions, definitions of the microstructural strain and stress, and the use of homogenization methods to establish the macro-micro relationship for granular materials.

The micromorphic theory [18-23] is a microstructural continuum approach developed from Mindlin-Eringen theory (linear elasticity [18] and micromorphic mechanics [19]) which has the capability to describe the deformation of complex microstructures. It assumes that the material is a deformable coordinator composed of macroscopic matter and microscopic matter, and the theory considers a material point as a volume cell. And degrees of freedom are introduced to describe the macro-micro deformation and rotation of microstructures. Based on this method, studies [24-28] have developed extended micromorphic theories including the micromorphic electromagnetic theory, the micromorphic thermoplasticity, the linear micromorphic elasticity, etc. And various micromorphic elastic and elastoplastic models

[29-36] are proposed to investigate the physical and mechanical properties influenced by microstructures for crystalline materials, heterogeneous materials, hierarchical materials, etc. The micromorphic theory has been applied to model granular materials. Misra and Poorsolhjouy [22,23,36,37] proposed a micromorphic model based on micromechanics and applied the model to investigate the dispersion behavior and predict the frequency band gap for granular elastic media [22]. Their model provides an approach to identify the complex macroscopic parameters of the micromorphic model based on micromechanics [22,37]. As mentioned above, Chang and Ma [4] also developed the continuum model based on micromechanics, and their model proposed an asymmetric macro strain based on micromechanics for granular materials and derived a macro stress $\sigma_{ij}$ as the form $\sigma_{ij} = \frac{1}{V}\sum f_i l_j$ conjugated with the macro strain based on the principle of energy equilibrium [15]. Thus, the asymmetric Cauchy stress is conjugated with the asymmetric strain [4]. However, there are some drawbacks in Misra-Poorsolhjouy micromorphic model [22,23,37]. In their model, a symmetric strain is proposed and a symmetric Cauchy stress should be obtained based on the conjugation of energy, but the model actually obtained an asymmetric Cauchy stress as $\sigma_{ij} = \frac{1}{V}\sum f_i l_j$ because of the effect of the contact couple [38]. So, the derivation in the model is imperfect, and this study checks the drawbacks of formulas in Misra-Poorsolhjouy model and proposes a new approach to avoid the drawbacks.

In this study, based on the micromorphic theory [18,19,22,23], a micromechanics-based micromorphic model is developed for granular materials. The material point is considered as a volume element of granular assembly. Different from the models of Mindlin-Eringen [18,19] and Misra-Poorsolhjouy [22,23,37], this study considers: (1) a symmetric Cauchy stress and a symmetric couple stress are obtained conjugating a symmetric strain and a symmetric curvature respectively. (2) The independent rotation of particle is introduced, and the relative displacement in a contact pair is influenced by the rotation of particle. (3) The model ignores the second-order micro-deformation gradient terms and obtains first-order constitutive relationships including the stress-strain relationship and the couple stress-curvature relationship. (4) Both the translation and the rotation of particle are decomposed. In the theory of Mindlin-Eringen [18,19], the macroscopic motion is a summation of the microscopic motion and a fluctuation between the macro-micro motion. However, in this study, the microscopic actual motion (translation and rotation) is regarded as a summation of an average of the microscopic actual motions and a fluctuation related to the average motion. A hypothesis is proposed that the macroscopic translation is considered as the average of the microscopic actual translations and the macroscopic rotation is the sum of the average of the microscopic actual rotations and the rigid rotation. Based on the decomposition proposed by this study, the microscopic constitutive relationships and microscopic deformation energy can be obtained. Combining the macroscopic deformation energy summed by microscopic deformation energy with the microscopic constitutive relationships, the macroscopic constitutive relationships and the corresponding constitutive moduli are derived as expressions of the microstructural information. In these macroscopic constitutive relationships, the Cauchy stress and the couple stress are symmetric conjugated with the symmetric strain and the symmetric curvature respectively, and the relative stress and the relative couple stress are asymmetric conjugated with the asymmetric relative strain and the asymmetric relative curvature respectively.

**2. Kinematics and decomposition for displacement and rotation**

According to the method of Mindlin [18], we define a globe coordinate system $x$ for a volume element $V$ representing a material point, and in $V$, a local coordinate system $X$ is defined where the origin is at the barycenter of $V$ and the axes are parallel to the $x$, as Figure 1 shows. Consider two particles $p$ and its neighboring $q$ in the volume element, and the displacement of $p$ is related to the displacement of $q$ by a first-order Taylor series expansion as

$$\phi_i^p = \phi_i^q + \phi_{i,j}^q l_j + \cdots \tag{1}$$

where $\phi_i$ is the microscopic displacement and it is a sum of products of specified functions of the $X$ and arbitrary functions of the $x$ and $t$ [18]. $l_j$ is a branch vector connecting the centroids of two particles in a contact pair. And a microscopic displacement gradient is defined as [18]

$$\psi_{ij} = \frac{\partial \phi_i}{\partial X_j} = \psi_{ij}(x,t) \tag{2}$$

so $\psi_{ij}$ is only a function of the $x$ and $t$.

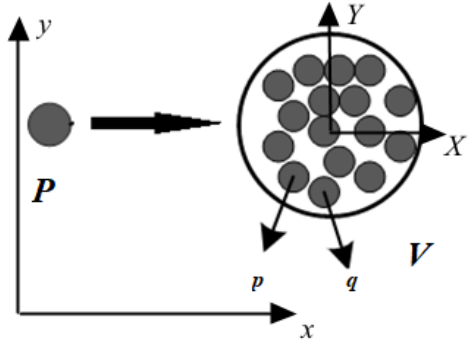

**FIGURE 1** Material point $P$ and its microstructural volume element $V$

We consider that the macroscopic displacement is considered as an average of the microscopic displacements, so the microscopic displacement can be regard as a summation of the macroscopic displacement and a fluctuation between the macro-micro displacement. Therefore, combining the relationship between the macroscopic and microscopic displacement gradient, a relative displacement gradient is defined by

$$\gamma_{ij} = \psi_{ij} - \bar{\phi}_{i,j} \tag{3}$$

where $\gamma_{ij}$ is regarded as a fluctuation of displacement gradient, $\bar{\phi}_{i,j}$ is the displacement gradient of macroscopic displacement $\bar{\phi}_i$, and we assume $\bar{\phi}_i = \bar{\phi}_i(x,t)$. Therefore, we define the macroscopic strain as

$$\varepsilon_{ij} = \bar{\phi}_{(i,j)} \equiv \frac{1}{2}(\bar{\phi}_{i,j} + \bar{\phi}_{j,i}) \tag{4}$$

where $\bar{\phi}_{(i,j)}$ is the symmetric part of $\bar{\phi}_{i,j}$.

Combining Equations 1 and 2, and considering the rotation of particles and ignoring the high-order terms, the relative displacement is obtained as [4]

$$\delta_i^{pq} = \phi_i^p - \phi_i^q + e_{ijk}\left(\chi_j^p r_k^p - \chi_j^q r_k^q\right) = \bar{\phi}_{i,j}l_j + \gamma_{ij}l_j + e_{ijk}\chi_{j,l}l_l r_k + e_{ijk}\chi_j l_k \qquad (5)$$

where $r_k$ is the radius of a reference particle, $e_{ijk}$ is the alternating tensor, $\chi_i$ is the particle rotation, and we consider that the degree of freedom for the rotation $\chi_i$ is independent of the displacement field. Then, the relative displacement is decomposed into three parts as

$$\delta_i^M = \bar{\phi}_{i,j}l_j, \ \delta_i^m = \gamma_{ij}l_j, \ \delta_i^{Rr} = e_{ijk}\chi_{j,l}l_l r_k + e_{ijk}\chi_j l_k \qquad (6)$$

where $\delta_i^M$ is the displacement caused by the macroscopic mean strain, $\delta_i^m$ is the displacement caused by the fluctuation of displacement gradient, $\delta_i^{Rr}$ is the displacement caused by the rotation of particle. Thus, the relative displacement between particles is comprised of the macroscopic displacement, the displacement of fluctuation and the displacement of rotation.

Similarly, the rotation of $p$ is expressed as the rotation of $q$ by the Taylor series expansion:

$$\chi_i^p = \chi_i^q + \chi_{i,j}^q l_j + \cdots \qquad (7)$$

$\chi_i$ is assumed to be a sum of products of specified functions of the $X$ and arbitrary functions of the $x$ and $t$, and the gradient of the rotation $\chi_i$ is obtained as

$$\varphi_{ij} = \frac{\partial \chi_i}{\partial X_j} = \varphi_{ij}(x,t) \qquad (8)$$

Then, ignoring the high-order term in Equation 7, the relative rotation between two particles is obtained by $\theta_i^{pq} = \varphi_{ij}l_j$. Taking a similar consideration into the decomposition for the rotation, the microscopic actual rotation of particle is decomposed into an average rotation and a fluctuation of rotation relative to the average rotation. And the gradient of the fluctuation of rotation, $\alpha_{ij}$, is expressed as

$$\alpha_{ij} = \varphi_{ij} - \bar{\chi}_{i,j} \qquad (9)$$

where $\bar{\chi}_i = \bar{\chi}_i(x,t)$ is the average rotation.
Thus, the relative rotation in Equation 8 is expressed as

$$\theta_i^{pq} = \varphi_{ij}l_j = \bar{\chi}_{i,j}l_j + \alpha_{ij}l_j \qquad (10)$$

Then, the relative rotation is decomposed into two parts as

$$\theta_i^R = \bar{\chi}_{i,j}l_j, \ \theta_i^r = \alpha_{ij}l_j \qquad (11)$$

And the components of $\delta_i^{Rr}$ in Equation 6 include

$$\delta_i^R = e_{ijk}\bar{\chi}_{j,l}l_l r_k, \ \delta_i^r = e_{ijk}\alpha_{jl}l_l r_k \qquad (12)$$

And the curvature is defined as the symmetrical part of the gradient of the macroscopic rotation referring to the modified couple stress theory [39]:

$$\kappa_{ij} = \bar{\chi}_{(i,j)} \equiv \frac{1}{2}(\bar{\chi}_{i,j} + \bar{\chi}_{j,i}) \qquad (13)$$

## 3. Macroscopic and microscopic constitutive relationships

We consider that the macroscopic deformation energy density $W$ is a function of four continuous variables, that is to say, $W = W(\varepsilon_{ij}, \gamma_{ij}, \kappa_{ij}, \alpha_{ij})$. Then, we obtain macroscopic stress measures by derivatives of the macroscopic deformation energy density $W$ as

$$\sigma_{ij} = \frac{\partial W}{\partial \varepsilon_{ij}}, \quad \tau_{ij} = \frac{\partial W}{\partial \gamma_{ij}}, \quad \mu_{ij} = \frac{\partial W}{\partial \kappa_{ij}}, \quad \nu_{ij} = \frac{\partial W}{\partial \alpha_{ij}} \tag{14}$$

where $\sigma_{ij}$ is the Cauthy stress, $\tau_{ij}$ is the relative stress, $\mu_{ij}$ is the couple stress, $\nu_{ij}$ is the relative couple stress. Because $\varepsilon_{ij}$ and $\kappa_{ij}$ are symmetric tensor, the Cauthy stress $\sigma_{ij}$ and the couple stress $\mu_{ij}$ should be symmetric based on the conjugation of energy. Tordesillas and Walsh [40] thought that the macroscopic deformation energy density is the summation of the microscopic deformation energy density between particles. And the microscopic deformation energy density, $w^c$, is defined by kinematic measures of a contact pair c as $w^c = w^c(\delta_i^M, \delta_i^m, \delta_i^R, \delta_i^r, \theta_i^R, \theta_i^r)$, and the macroscopic energy density is expressed as the volume average of the microscopic deformation energy density:

$$W = \frac{1}{V} \sum_{c=1}^{n} w^c(\delta_i^M, \delta_i^m, \delta_i^R, \delta_i^r, \theta_i^R, \theta_i^r) \tag{15}$$

Then, taking derivatives for the microscopic energy density, the inter-particle contact force $f_i^c$ and contact moment $m_i^c$ are obtained by

$$\frac{\partial w^c}{\partial \delta_i^{c\xi}} = f_i^{c\xi}, \quad \frac{\partial w^c}{\partial \theta_i^{c\eta}} = m_i^{c\eta} \tag{16}$$

where $\xi = M, m, R$, $\eta = R, r$.

Substituting Equations 15 and 16 into Equation 14, macroscopic stress measures can be obtained by inter-particle contact force and contact moment, where the Cauchy stress is expressed as

$$\sigma_{ij} = \frac{\partial W}{\partial \varepsilon_{ij}} = \frac{1}{V} \sum_{c=1}^{n} \frac{\partial w^c}{\partial \delta_k^{cM}} \frac{\partial \delta_k^{cM}}{\partial \varepsilon_{ij}} = \frac{1}{V} \sum_{c=1}^{n} \frac{\partial w^c}{\partial \delta_k^{cM}} \frac{\partial \bar{\phi}_{k,l} l_l^c}{\partial \varepsilon_{ij}} \tag{17}$$

For convenience in the further derivation, in Equation 17, the derivative $\frac{\partial \bar{\phi}_{k,l}}{\partial \varepsilon_{ij}}$ should be obtained. Note that

$$\frac{\partial \bar{\phi}_{k,l}}{\partial \varepsilon_{ij}} = \frac{\partial \bar{\phi}_{l,k}}{\partial \varepsilon_{ij}} \tag{18}$$

Then, the derivative $\frac{\partial \bar{\phi}_{k,l}}{\partial \varepsilon_{ij}}$ is given by

$$\frac{\partial \bar{\phi}_{k,l}}{\partial \varepsilon_{ij}} = \frac{1}{2} \frac{\partial \bar{\phi}_{k,l} + \partial \bar{\phi}_{l,k}}{\partial \varepsilon_{ij}} = \frac{1}{2} \frac{2 \partial \varepsilon_{kl}}{\partial \varepsilon_{ij}} = \frac{1}{2} \frac{\partial \varepsilon_{kl} + \partial \varepsilon_{lk}}{\partial \varepsilon_{ij}} = \frac{1}{2} (\delta_{ik} \delta_{jl} + \delta_{il} \delta_{jk}) \tag{19}$$

Thus, the Cauchy stress in Equation 17 is obtained as

$$\sigma_{ij} = \frac{\partial W}{\partial \varepsilon_{ij}} = \frac{1}{V} \sum_{c=1}^{n} \frac{\partial w^c}{\partial \delta_k^{cM}} \frac{\partial \delta_k^{cM}}{\partial \varepsilon_{ij}} = \frac{1}{V} \sum_{c=1}^{n} \frac{\partial w^c}{\partial \delta_k^{cM}} \frac{\partial \bar{\phi}_{k,l} l_l^c}{\partial \varepsilon_{ij}}$$

$$= \frac{1}{V} \sum_{c=1}^{n} f_k^{cM} \frac{1}{2} (\delta_{ik} \delta_{jl} + \delta_{il} \delta_{jk}) l_l^c = \frac{1}{V} \sum_{c=1}^{n} \frac{1}{2} (f_i^{cM} l_j^c + f_j^{cM} l_i^c) \tag{20}$$

Therefore, the Cauchy stress $\sigma_{ij}$ in Equation 20 is symmetric, while the Cauchy stress $\sigma_{ij} = \frac{1}{V}\sum f_i l_j$ is not necessarily symmetric in Misra-Poorsolhjouy model [22,23,37]. And the couple stress is expressed as

$$\mu_{ij} = \frac{\partial W}{\partial \kappa_{ij}} = \frac{1}{V}\left(\sum_{c=1}^{n}\frac{\partial w^c}{\partial \delta_l^{cR}}\frac{\partial \delta_l^{cR}}{\partial \kappa_{ij}} + \sum_{c=1}^{n}\frac{\partial w^c}{\partial \theta_l^{cR}}\frac{\partial \theta_l^{cR}}{\partial \kappa_{ij}}\right) \tag{21}$$

where the derivatives $\frac{\partial \delta_l^{cR}}{\partial \kappa_{ij}}$ and $\frac{\partial \theta_l^{cR}}{\partial \kappa_{ij}}$ are derived as

$$\frac{\partial \delta_l^{cR}}{\partial \kappa_{ij}} = \frac{\partial \bar{\chi}_{m,n} e_{lmk} l_n^c r_k^c}{\partial \kappa_{ij}} = e_{lmk} l_n^c r_k^c \frac{\partial \bar{\chi}_{m,n}}{\partial \kappa_{ij}} = \frac{1}{2}(\delta_{im}\delta_{jn} + \delta_{in}\delta_{jm})e_{lmk} l_n^c r_k^c \tag{22}$$

$$\frac{\partial \theta_l^{cR}}{\partial \kappa_{ij}} = \frac{\partial \bar{\chi}_{l,n} l_n^c}{\partial \kappa_{ij}} = \frac{1}{2}\frac{\partial \bar{\chi}_{l,n} + \partial \bar{\chi}_{n,l}}{\partial \kappa_{ij}} l_n^c = \frac{1}{2}\frac{\partial \kappa_{ln} + \partial \kappa_{nl}}{\partial \kappa_{ij}} l_n^c = \frac{1}{2}(\delta_{il}\delta_{jn} + \delta_{in}\delta_{jl})l_n^c \tag{23}$$

Therefore, the couple stress in Equation 21 is obtained as

$$\mu_{ij} = \frac{\partial W}{\partial \kappa_{ij}} = \frac{1}{V}\left(\sum_{c=1}^{n}\frac{\partial w^c}{\partial \delta_l^{cR}}\frac{\partial \delta_l^{cR}}{\partial \kappa_{ij}} + \sum_{c=1}^{n}\frac{\partial w^c}{\partial \theta_l^{cR}}\frac{\partial \theta_l^{cR}}{\partial \kappa_{ij}}\right)$$

$$= \frac{1}{V}\left(\sum_{c=1}^{n} f_l^{cR}\frac{1}{2}(\delta_{im}\delta_{jn} + \delta_{in}\delta_{jm})e_{lmk} l_n^c r_k^c + \sum_{c=1}^{n} m_l^{cR}\frac{1}{2}(\delta_{il}\delta_{jn} + \delta_{in}\delta_{jl})l_n^c\right)$$

$$= \frac{1}{V}\sum_{c=1}^{n}\frac{1}{2}f_l^{cR}(e_{lik} l_j^c + e_{ljk} l_i^c)r_k^c + \frac{1}{V}\sum_{c=1}^{n}\frac{1}{2}(m_i^{cR} l_j^c + m_j^{cR} l_i^c) \tag{24}$$

As a result, this study obtains a symmetric couple stress which should be conjugated with a symmetric curvature based on the conjugation of energy. Then, a same operation leads to relative stress measure expressions which have the similar formats as the stress measure expressions:

$$\tau_{ij} = \frac{\partial W}{\partial \gamma_{ij}} = \frac{1}{V}\sum_{c=1}^{n}\frac{\partial w^c}{\partial \delta_k^{cm}}\frac{\partial \delta_k^{cm}}{\partial \gamma_{ij}} = \frac{1}{V}\sum_{c=1}^{n} f_k^{cm}\frac{\partial \gamma_{kl}}{\partial \gamma_{ij}}l_l^c = \frac{1}{V}\sum f_i^{cm} l_j^c \tag{25}$$

$$\nu_{ij} = \frac{\partial W}{\partial \alpha_{ij}} = \frac{1}{V}\left(\sum_{c=1}^{n}\frac{\partial w^c}{\partial \delta_l^{cr}}\frac{\partial \delta_l^{cr}}{\partial \alpha_{ij}} + \frac{\partial w^c}{\partial \theta_k^{cr}}\frac{\partial \theta_k^{cr}}{\partial \alpha_{ij}}\right)$$

$$= \frac{1}{V}\sum f_l^{cr}\frac{\partial \alpha_{mn} e_{lmk} l_n^c r_k^c}{\partial \alpha_{ij}} + \frac{1}{V}\sum_{c=1}^{n} m_k^{cr}\frac{\partial \alpha_{kl}}{\partial \alpha_{ij}}l_l^c = \frac{1}{V}\sum_{c=1}^{n} e_{ikl} f_l^{cr} l_j^c r_k^c + \frac{1}{V}\sum m_i^{cr} l_j^c \tag{26}$$

Therefore, macroscopic stress measures in Equation 20 and Equations 24-26 are obtained by inter-particle contact force $f_i^c$ and contact moment $m_i^c$ and the branch vector $l_j^c$ representing the microstructural information. The Cauchy stress and the couple stress are symmetric, while the relative stress measures are asymmetric.

To develop the microscopic constitutive relationships for particles, we first define a local coordinate system for each contact as Figure 2 [41] shows, where **n** is a unit vector normal to the contact plane, and the other two orthogonal unit vectors, **s** and **t**, are on the contact plane by Hicher and Chang [41]

$$\mathbf{n} = \cos\alpha\, \mathbf{e}_1 + \sin\alpha \cos\beta\, \mathbf{e}_2 + \sin\alpha \sin\beta\, \mathbf{e}_3 \tag{27}$$

$$\mathbf{s} = \frac{d\mathbf{n}}{d\alpha} = -\sin\alpha\, \mathbf{e}_1 + \cos\alpha \cos\beta\, \mathbf{e}_2 + \cos\alpha \sin\beta\, \mathbf{e}_3 \tag{28}$$

$$\mathbf{t} = \mathbf{n} \times \mathbf{s} = -\sin\beta\, \mathbf{e}_2 + \cos\beta\, \mathbf{e}_3 \tag{29}$$

where $\mathbf{e}_1, \mathbf{e}_2, \mathbf{e}_3$ are the unit vectors parallel to the axis of coordinates.

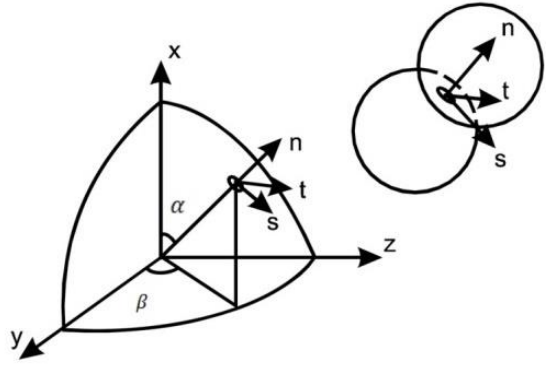

**FIGURE 2** Local coordinate system at a contact of particles [41]

Then, ignoring the cross terms, the microscopic deformation energy density is expressed as

$$w^c = \sum_c f_n^{c\xi} \delta_n^{c\xi} + f_s^{c\xi} \delta_s^{c\xi} + f_t^{c\xi} \delta_t^{c\xi} + m_n^{c\eta} \theta_n^{c\eta} + m_s^{c\eta} \theta_s^{c\eta} + m_t^{c\eta} \theta_t^{c\eta} \tag{30}$$

where $\xi = M, m, R$, $\eta = R, r$. Subscripts $n, s, t$ donates components at axis direction of the above local coordinate system in Figure 2. Furthermore, for the linear isotropic elasticity, the microscopic deformation energy density is expressed as

$$w^c = \frac{1}{2}\left[\sum_c K_n^\xi \left(\delta_n^{c\xi}\right)^2 + K_s^\xi \left(\delta_s^{c\xi}\right)^2 + K_t^\xi \left(\delta_t^{c\xi}\right)^2 + G_n^\eta \left(\theta_n^{c\eta}\right)^2 + G_s^\eta \left(\theta_s^{c\eta}\right)^2 + G_t^\eta \left(\theta_t^{c\eta}\right)^2\right] \tag{31}$$

where $K_n^\xi$, $K_s^\xi$, $K_t^\xi$ and $G_n^\eta$, $G_s^\eta$, $G_t^\eta$ are the contact stiffness parameters for contact forces and contact moments. For two particles at a contact, the contact area is circle, so the contact stiffness parameters are same at **s** and **t** directions, and we define that

$$K_s^\xi = K_t^\xi = K_w^\xi, \quad G_s^\eta = G_t^\eta = G_w^\eta \tag{32}$$

Therefore, microscopic constitutive relationships and stiffness matrices are written as the following forms by relating the local coordinates to the global coordinates:

$$f_i^{c\xi} = K_{ij}^\xi \delta_j^{c\xi}, \quad K_{ij}^\xi = K_n^\xi n_i n_j + K_w^\xi \left(s_i s_j + t_i t_j\right) \tag{33}$$

$$m_i^{c\eta} = G_{ij}^\eta \theta_j^{c\eta}, \quad G_{ij}^\eta = G_n^\eta n_i n_j + G_w^\eta \left(s_i s_j + t_i t_j\right) \tag{34}$$

where $\xi = M, m, R$, $\eta = R, s, u$, $i, j = 1,2,3$. Then substituting Equations 33 and 34 into Equation 20 and Equations 24-26, we can obtain the Cauchy stress and the couple stress conjugated with the symmetric strain and the symmetric curvature respectively, and obtain the relative stress and the relative couple stress conjugated with the asymmetric relative strain and the asymmetric relative curvature respectively. The Cauchy stress is expressed as

$$\sigma_{ij} = \frac{1}{V}\sum_{c=1}^{n}\frac{1}{2}(f_i^{cM}l_j^c + f_j^{cM}l_i^c) = \frac{1}{V}\sum_{c=1}^{n}\frac{1}{2}(K_{ik}^M\delta_k^{cM}l_j^c + K_{jl}^M\delta_l^{cM}l_i^c)$$

$$= \frac{1}{V}\sum_{c=1}^{n}\frac{1}{2}(K_{ik}^M\bar{\phi}_{k,l}l_i^c l_j^c + K_{jl}^M\bar{\phi}_{l,k}l_k^c l_i^c) = \frac{1}{V}\sum_{c=1}^{n}\frac{1}{2}(K_{kl}^M\bar{\phi}_{k,l}l_i^c l_j^c + K_{lk}^M\bar{\phi}_{l,k}l_j^c l_i^c)$$

$$= \frac{1}{V}\sum_{c=1}^{n}\frac{1}{2}K_{kl}^M(\bar{\phi}_{k,l} + \bar{\phi}_{l,k})l_i^c l_j^c = \frac{1}{V}\sum_{c=1}^{n}K_{kl}^M l_i^c l_j^c \varepsilon_{kl} = \left(\frac{1}{V}\sum_{c=1}^{n}K_{kl}^M l_i^c l_j^c\right)\varepsilon_{kl} = C_{ijkl}^M \varepsilon_{kl}$$

(35)

And the terms of the couple stress in Equation 24 is derived as following:

$$f_l^{cR} e_{lik} l_j^c r_k^c = K_{lm}^R \delta_m^{cR} e_{lik} l_j^c r_k^c = K_{nm}^R e_{mkq} \bar{\chi}_{k,l} l_i^c r_q^c e_{nip} l_j^c r_p^c$$

$$= K_{nm}^R e_{mkq} \bar{\chi}_{k,l} (l_i^c \delta_{li} e_{nip} l_j^c) r_p^c r_q^c = K_{nm}^R e_{mkq} \bar{\chi}_{k,l} (l_i^c e_{nlp} l_j^c) r_p^c r_q^c$$

$$= (K_{nm}^R e_{mkq} e_{nlp} l_i^c l_j^c r_p^c r_q^c) \bar{\chi}_{k,l}$$

(36a)

$$f_l^{cR} e_{ljk} l_i^c r_k^c = K_{ln}^R \delta_n^{cR} e_{ljk} l_i^c r_k^c = K_{mn}^R e_{nlq} \bar{\chi}_{l,k} l_k^c r_q^c e_{mjp} l_i^c r_p^c$$

$$= K_{mn}^R e_{nlq} \bar{\chi}_{l,k} (l_j^c \delta_{kj} e_{mjp} l_i^c) r_p^c r_q^c = K_{mn}^R e_{nlq} \bar{\chi}_{l,k} (l_j^c e_{mkp} l_i^c) r_p^c r_q^c$$

$$= (K_{mn}^R e_{nlq} e_{mkp} l_j^c l_i^c r_p^c r_q^c) \bar{\chi}_{l,k}$$

(36b)

Therefore, the first term of Equation 24 is obtained as the sum of Equations 36a and 36b:

$$f_l^{cR} e_{lik} l_j^c r_k^c + f_l^{cR} e_{ljk} l_i^c r_k^c = (K_{nm}^R e_{mkq} e_{nlp} l_i^c l_j^c r_p^c r_q^c)(\bar{\chi}_{k,l} + \bar{\chi}_{l,k})$$

$$= 2(K_{nm}^R e_{mkq} e_{nlp} l_i^c l_j^c r_p^c r_q^c)\kappa_{kl}$$

(37)

And $m_i^{cR} l_j^c + m_j^{cR} l_i^c$ can be obtained in the same manner with Equation 35. Therefore, the couple stress in Equation 24 is expressed as

$$\mu_{ij} = \left(\frac{1}{V}\sum_{c=1}^{n} K_{mn}^R e_{nlq} e_{mkp} l_i^c l_j^c r_p^c r_q^c\right)\kappa_{kl} + \left(\frac{1}{V}\sum_{c=1}^{n} G_{kl}^R l_i^c l_j^c\right)\kappa_{kl} = (B_{ijkl}^R + C_{ijkl}^R)\kappa_{kl}$$

(38)

Then, the relative stress and the relative couple stress conjugated with the relative strain and the relative curvature respectively are derived as

$$\tau_{ij} = \frac{1}{V}\sum_{c=1}^{n} f_i^{cm} l_j^c = \left(\frac{1}{V}\sum_{c=1}^{n} K_{kl}^m l_i^c l_j^c\right)\gamma_{kl} = C_{ijkl}^m \gamma_{kl}$$

$$v_{ij} = \frac{1}{V}\sum_{c=1}^{n} e_{ikl}f_l^{cr}l_j^c r_k^c + \frac{1}{V}\sum_{c=1}^{n} m_i^{cr}l_j \tag{39}$$

$$= \left(\frac{1}{V}\sum_{c=1}^{n} K_{mn}^R e_{nlq}e_{mkp}l_i^c l_j^c r_p^c r_q^c\right)\alpha_{kl} + \left(\frac{1}{V}\sum_{c=1}^{n} G_{kl}^r l_i^c l_j^c\right)\alpha_{kl} = (B_{ijkl}^r + C_{ijkl}^r)\alpha_{kl} \tag{40}$$

From Equation 35 and Equations 38 to 40, macroscopic constitutive relationships are first-order, because of the introduce of the rotation of particle instead of the micro-deformation gradient in the model of Mindlin-Eringen [18,19]. And macroscopic constitutive parameters are functions of contact stiffness parameters and internal lengths. For a volume element, contact stiffness parameters and internal lengths are usually different among contacts. To simply the computation, we assume that the material is isotropic, and Chang and Ma [42] proposed a directional density distribution function $\xi(\alpha,\beta) = 1/4\pi$. Therefore, the macroscopic constitutive parameters are written as integral forms from discrete summation forms:

$$C_{ijkl}^M = \frac{1}{V}\sum_{c=1}^{n} K_{kl}^M l_i^c l_j^c = l^2 N_V \int_0^{\pi}\int_0^{2\pi} \left(K_{kl}^M n_i n_j\right)\xi \sin\alpha \, d\beta \, d\alpha \tag{41}$$

$$C_{ijkl}^m = \frac{1}{V}\sum_{c=1}^{n} K_{kl}^m l_i^c l_j^c = l^2 N_V \int_0^{\pi}\int_0^{2\pi} \left(K_{kl}^m n_i n_j\right)\xi \sin\alpha \, d\beta \, d\alpha \tag{42}$$

$$B_{ijkl}^R = \frac{1}{V}\sum_{c=1}^{n} K_{mn}^R e_{nlq}e_{mkp}l_i^c l_j^c r_p^c r_q^c$$

$$= l^2 r^2 N_V \int_0^{\pi}\int_0^{2\pi} \left(K_{mn}^R e_{nlq}e_{mkp}n_i n_j n_p n_q\right)\xi \sin\alpha \, d\beta \, d\alpha \tag{43}$$

$$B_{ijkl}^r = \frac{1}{V}\sum_{c=1}^{n} K_{mn}^R e_{nlq}e_{mkp}l_i^c l_j^c r_p^c r_q^c$$

$$= l^2 r^2 N_V \int_0^{\pi}\int_0^{2\pi} \left(K_{mn}^r e_{nlq}e_{mkp}n_i n_j n_p n_q\right)\xi \sin\alpha \, d\beta \, d\alpha \tag{44}$$

$$C_{ijkl}^R \frac{1}{V}\sum_{c=1}^{n} G_{kl}^R l_i^c l_j^c = l^2 N_V \int_0^{\pi}\int_0^{2\pi} \left(G_{kl}^R n_i n_j\right)\xi \sin\alpha \, d\beta \, d\alpha \tag{45}$$

$$C_{ijkl}^r = \frac{1}{V}\sum_{c=1}^{n} G_{kl}^r l_i^c l_j^c = l^2 N_V \int_0^{\pi}\int_0^{2\pi} \left(G_{kl}^r n_i n_j\right)\xi \sin\alpha \, d\beta \, d\alpha$$

(46)

where $N_V$ represents the volume density of the volume element. Thus, solving these above integrals, we obtain

$$C_{iiii}^{\xi} = \frac{l^2 N_V}{15}\left(3K_n^{\xi} + 2K_w^{\xi}\right), \quad C_{ijij}^{\xi} = \frac{l^2 N_V}{15}\left(K_n^{\xi} - K_w^{\xi}\right),$$

$$C_{iijj}^{\xi} = \frac{l^2 N_V}{15}\left(K_n^{\xi} + 4K_w^{\xi}\right), \quad C_{ijji}^{\xi} = \frac{l^2 N_V}{15}\left(K_n^{\xi} - K_w^{\xi}\right),$$

$$C_{ijkl}^{\xi} = 0, \text{ otherwise} \tag{47}$$

$$C_{iiii}^{\eta} = \frac{l^2 N_V}{15}\left(3G_n^{\eta} + 2G_w^{\eta}\right), \quad C_{ijij}^{\eta} = \frac{l^2 N_V}{15}\left(G_n^{\eta} - G_w^{\eta}\right),$$

$$C_{iijj}^{\eta} = \frac{l^2 N_V}{15}\left(G_n^{\eta} + 4G_w^{\eta}\right), \quad C_{ijji}^{\eta} = \frac{l^2 N_V}{15}\left(G_n^{\eta} - G_w^{\eta}\right),$$

$$C_{ijkl}^{\eta} = 0, \text{ otherwise} \tag{48}$$

$$B_{iiii}^{\eta} = \frac{l^2 r^2 N_V}{15}\left(2K_w^{\eta}\right), \quad B_{ijij}^{\eta} = \frac{l^2 r^2 N_V}{15}\left(-K_w^{\eta}\right),$$

$$B_{iijj}^{\eta} = \frac{l^2 r^2 N_V}{15}\left(4K_w^{\eta}\right), \quad B_{ijji}^{\eta} = \frac{l^2 r^2 N_V}{15}\left(-K_w^{\eta}\right)$$

$$B_{ijkl}^{\eta} = 0, \text{ otherwise} \tag{49}$$

where $\xi = M, m, \eta = R, r$. $i, j = 1,2,3$, and $i \neq j$.

## 4 Balance equations and boundary conditions

Based on the principle of virtual work, the stress measures ($\sigma_{ij}$, $\tau_{ij}$, $\mu_{ij}$, $\nu_{ij}$) are respectively dual in energy to the deformation measures ($\varepsilon_{ij}, \gamma_{ij}, \kappa_{ij}, \alpha_{ij}$). Therefore, the principle of virtual work in a volume element $V$ with the boundary $S$ is written as

$$\delta \int_0^t (\mathcal{T} - \mathcal{W})\mathrm{d}t + \int_0^t \delta \mathcal{W}^{ext}\mathrm{d}t = 0$$

(50)

where $\delta W$ is the variation of the potential energy density $W$ and should be written as

$$\delta W = \sigma_{ij}\delta\varepsilon_{ij} + \tau_{ij}\delta\gamma_{ij} + \mu_{ij}\delta\kappa_{ij} + \nu_{ij}\delta\alpha_{ij}$$
$$= \sigma_{ij}\delta\bar{u}_{i,j} + \tau_{ij}(\delta u_{i,j} - \delta\bar{u}_{i,j}) + \mu_{ij}\delta\bar{\chi}_{i,j} + \nu_{ij}(\delta\chi_{i,j} - \delta\bar{\chi}_{i,j})$$
$$= (\sigma_{ij} - \tau_{ij})\delta\bar{u}_{i,j} + \tau_{ij}\delta\psi_{ij} + (\mu_{ij} - \nu_{ij})\delta\bar{\chi}_{i,j} + \nu_{ij}\delta\varphi_{ij}$$
$$= \left[(\sigma_{ij} - \tau_{ij})\delta\bar{u}_i\right]_{,j} - (\sigma_{ij} - \tau_{ij})_{,j}\delta\bar{u}_i + \tau_{ij}\delta\psi_{ij} + \left[(\mu_{ij} - \nu_{ij})\delta\bar{\chi}_i\right]_{,j} - (\mu_{ij} - \nu_{ij})_{,j}\delta\bar{\chi}_i$$
$$+ \nu_{ij}\delta\varphi_{ij}$$

(51)

The macroscopic rotation can be obtained as the sum of the rigid rotation and the average of microscopic rotations of particles:

$$\Gamma_i = e_{ijk}\bar{u}_{k,j} + \bar{\chi}_i \tag{52}$$

Then, the terms in Equation 51 are derived as

$$[(\mu_{ij} - \nu_{ij})\delta\bar{\chi}_i]_{,j} - (\mu_{ij} - \nu_{ij})_{,j}\delta\bar{\chi}_i$$

$$= [(\mu_{ij} - \nu_{ij})\delta\Gamma_i]_{,j} - [(\mu_{ij} - \nu_{ij})\delta e_{ikl}\bar{u}_{l,k}]_{,j} - (\mu_{ij} - \nu_{ij})_{,j}\delta\Gamma_i + (\mu_{ij} - \nu_{ij})_{,j}\delta e_{ikl}\bar{u}_{l,k}$$

$$= [(\mu_{ij} - \nu_{ij})\delta\Gamma_i]_{,j} - [e_{ikl}(\mu_{ij} - \nu_{ij})\delta\bar{u}_l]_{,kj} + [e_{ikl}(\mu_{ij} - \nu_{ij})_{,k}\delta\bar{u}_l]_{,j} - (\mu_{ij} - \nu_{ij})_{,j}\delta\Gamma_i$$

$$+ [e_{ikl}(\mu_{ij} - \nu_{ij})_{,j}\delta\bar{u}_l]_{,k} - [e_{ikl}(\mu_{ij} - \nu_{ij})_{,j}]_{,k}\delta\bar{u}_l$$

$$= [(\mu_{ij} - \nu_{ij})\delta\Gamma_i]_{,j} - [e_{lki}(\mu_{lj} - \nu_{lj})\delta\bar{u}_i]_{,kj} + [e_{lki}(\mu_{lj} - \nu_{lj})_{,k}\delta\bar{u}_i]_{,j} - (\mu_{ij} - \nu_{ij})_{,j}\delta\Gamma_i$$

$$+ [e_{lki}(\mu_{lj} - \nu_{lj})_{,j}\delta\bar{u}_i]_{,k} - e_{lki}(\mu_{lj} - \nu_{lj})_{,jk}\delta\bar{u}_i$$

(53)

$$\delta W = \int_V \left[-(\sigma_{ij} - \tau_{ij})_{,j} - e_{lki}(\mu_{lj} - \nu_{lj})_{,jk}\right]\delta\bar{u}_i dV + \int_V [-(\mu_{ij} - \nu_{ij})_{,j}]\delta\Gamma_i dV$$

$$+ \int_V \tau_{ij}\delta\psi_{ij} dV + \int_V \nu_{ij}\delta\varphi_{ij} dV$$

$$+ \int_s \left[(\sigma_{ij} - \tau_{ij})n_j + e_{lki}(\mu_{lj} - \nu_{lj})_{,k}n_j + e_{lki}(\mu_{lj} - \nu_{lj})_{,j}n_k\right]\delta\bar{u}_i dS$$

$$+ \int_s [(\mu_{ij} - \nu_{ij})n_j]\delta\Gamma_i dS - \int_s [e_{lki}(\mu_{lj} - \nu_{lj})\delta\bar{u}_i]_{,k} n_j dS$$

$$= \int_V \left[-(\sigma_{ij} - \tau_{ij})_{,j} - e_{lki}(\mu_{lj} - \nu_{lj})_{,jk}\right]\delta\bar{u}_i dV + \int_V [-(\mu_{ij} - \nu_{ij})_{,j}]\delta\Gamma_i dV + \int_V \tau_{ij}\delta\psi_{ij} dV$$

$$+ \int_V \nu_{ij}\delta\varphi_{ij} dV + \int_s \left[(\sigma_{ij} - \tau_{ij})n_j + e_{lki}(\mu_{lj} - \nu_{lj})_{,j}n_k\right]\delta\bar{u}_i dS$$

$$+ \int_s [(\mu_{ij} - \nu_{ij})n_j]\delta\Gamma_i dS - \int_s e_{lki}(\mu_{lj} - \nu_{lj})n_j\delta\bar{u}_{i,k} dS$$

(54)

And $\delta W^{ext}$ is the variation of the work done by external forces which is written as

$$\delta W^{ext} = \int_V f_i\delta\bar{u}_i dV + \int_V m_i\delta\Gamma_i dV + \int_V \Phi_{ij}\delta\psi_{ij} dV + \int_V \Pi_{ij}\delta\varphi_{ij} dV$$
$$+ \int_s t_i\delta\bar{u}_i dS + \int_s M_i\delta\Gamma_i dS$$

(55)

where $f_i$, $m_i$, $\Phi_{ij}$ respectively are the body force, moment and double force per unit volume, $t_i$, $M_i$, $T_{ij}$ respectively are the traction, moment and double traction per unit area.

Meanwhile, the variation of the total kinetic energy is written as

$$\delta \int_0^t \mathcal{T} dt = -\int_0^t dt \int_V \rho\ddot{\bar{u}}_i\delta\bar{u}_i + \rho'I'\ddot{\psi}_{ij}\delta\psi_{ij} + \rho I\ddot{\Gamma}_i\delta\Gamma_i + \rho'J'\ddot{\varphi}_{ij}\delta\varphi_{ij} dV$$

(56)

where $\rho$ and $\rho'$ are the macroscopic mass density and the microscopic mass density, $I$, $I'$, $J'$ are moments of inertia of an area expressed as $I = \frac{2}{3}r^2$, $I' = \frac{3}{5}r^2$, $J' = \frac{1}{2}r^4$ and $d$ is the length of a particle cell [18] as shown in Figure 3.

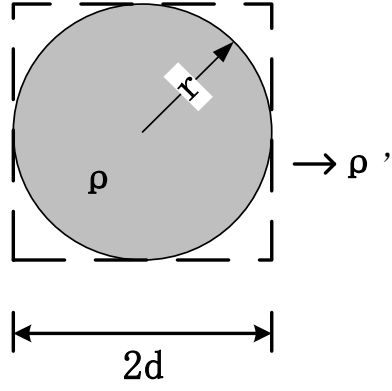

**FIGURE 3** The diagram of a particle cell

Then, substituting Equations 54-56 into Equation 50, the following equation can be obtained:

$$\int_V \left[ (\sigma_{ij} - \tau_{ij})_{,j} + e_{lki}(\mu_{lj} - \nu_{lj})_{,jk} + f_i - \rho \ddot{\bar{u}}_i \right] \delta \bar{u}_i \mathrm{d}V$$
$$+ \int_V [(\mu_{ij} - \nu_{ij})_{,j} + m_i - \rho I \ddot{\Gamma}_i] \delta \Gamma_i \mathrm{d}V$$
$$+ \int_V [-\tau_{ij} + \Phi_{ij} - \rho' I' \ddot{\psi}_{ij}] \delta \psi_{ij} \mathrm{d}V + \int_V [-\nu_{ij} + \Pi_{ij} - \rho' J' \ddot{\varphi}_{ij}] \delta \varphi_{ij} \mathrm{d}V$$
$$+ \int_s \left[ t_i - (\sigma_{ij} - \tau_{ij}) n_j - e_{lki}(\mu_{lj} - \nu_{lj})_{,j} n_k \right] \delta \bar{u}_i \mathrm{d}S + \int_s [M_i - (\mu_{ij} - \nu_{ij}) n_j] \delta \Gamma_i \mathrm{d}S$$
$$+ \int_s [e_{lki}(\mu_{lj} - \nu_{lj}) n_j] \delta \bar{u}_{i,k} \mathrm{d}S = 0$$

(57)

To satisfy Equation 50 all the time, it requires that every integral of Equation 57 should be zero, therefore, we can obtain four sets of balance equations:

$$\begin{cases} (\sigma_{ij} - \tau_{ij})_{,j} + e_{ilk}(\mu_{lj} - \nu_{lj})_{,jk} + f_i = \rho \ddot{\bar{u}}_i \\ (\mu_{ij} - \nu_{ij})_{,j} + m_i = \rho I \ddot{\Gamma}_i \\ -\tau_{ij} + \Phi_{ij} = \rho' I' \ddot{\psi}_{ij} \\ -\nu_{ij} + \Pi_{ij} = \rho' J' \ddot{\varphi}_{ij} \end{cases}$$

(58)

and three sets of boundary conditions:

$$\begin{cases} (\sigma_{ij} - \tau_{ij}) n_j + e_{ilk}(\mu_{lj} - \nu_{lj})_{,j} n_k = t_i \\ (\mu_{ij} - \nu_{ij}) n_j = M_i \\ e_{ijl}(\mu_{lk} - \nu_{lk}) n_k = 0 \end{cases}$$

(59)

Owning to the introduce of rotation of particle, this modified model obtains different balance equations and different boundary conditions from those in models of Mindlin-Eringen [18,19] and Misra-Poorsolhjouy [22].

## 7 Conclusions

The micromechanical approach is applied to develop a modified micromorphic continuum model for granular materials in this study, which can describe the microstructural information such as contact stiffness parameters and the internal lengths. Based on this approach, a symmetric Cauchy stress and a symmetric couple stress are obtained conjugating a symmetric strain and a symmetric curvature respectively, and asymmetric relative stress measures are obtained conjugating asymmetric relative strain measures. In this modified model, the independent rotation of particle is introduced, and the translation and the rotation of particles are decomposed, leading to a complete pattern of deformation. A hypothesis is proposed that the microscopic actual motion (translation and rotation) is regarded as a summation of an average of the microscopic actual motions and a fluctuation related the average motion. Based on the decomposition, the macroscopic constitutive relationships and the deformation moduli are derived based on the microstructural information. Because of the introduce of the independent rotation of particle, the micro-deformation gradient is ignored leading to the first-order macroscopic constitutive relationships. These constitutive relationships include the stress-strain relationship, the couple stress-curvature relationship and the corresponding relative stress measure-relative strain measure relationship. In the constitutive relationships, the Cauchy stress and the couple stress are symmetric conjugated with the symmetric strain and the symmetric curvature respectively, and the relative stress and the relative couple stress are asymmetric conjugated with the asymmetric relative strain and the asymmetric relative curvature respectively.

## Acknowledgement

The authors are pleased to acknowledge the support of this work by the National Natural Science Foundation of China through contract/grant number 11772237 and 11472196.